\documentclass[preprint,showpacs,preprintnumbers,amsmath,amssymb]{revtex4}

\usepackage[dvips]{graphicx}
\usepackage{graphicx}
\usepackage{dcolumn}
\usepackage{bm}
\usepackage{amsmath}
\usepackage{amsfonts}
\usepackage{amssymb}
\usepackage{color}
\usepackage{pstcol}
\usepackage[all]{xy}
\def\be{\begin{equation}}
\def\ee{\end{equation}}
\begin{document}
\title{Gowdy $T^3$ Cosmological Models in $N=1$ Supergravity}

\author{Alfredo Mac\'{\i}as}
 \email{amac@xanum.uam.mx}
\affiliation{Departamento de F\'{\i}sica, Universidad Autonoma Metropolitana--Iztapalapa,\\
A.P. 55--534, M\'exico D.F. 09340, M\'exico}
\author{Hernando Quevedo}
 \email{quevedo@nuclecu.unam.mx}
\affiliation{Instituto de Ciencias Nucleares,
Universidad Nacional Aut\'onoma de M\'exico\\
A.P. 70-543,  M\'exico D.F. 04510, M\'exico}
\author{Alberto S\'anchez}
 \email{asan@xanum.uam.mx}
 \affiliation{Departamento de F\'{\i}sica, Universidad Autonoma Metropolitana--Iztapalapa,\\
A. P. 55--534, M\'exico D.F. 09340, M\'exico}

\date{\today}

\begin{abstract}
We investigate the canonical quantization of supergravity $N=1$
in the case of a midisuperspace described by Gowdy $T^3$
cosmological models. The quantum constraints are analyzed
and the wave function of the universe is derived explicitly.
Unlike the minisuperspace case, we show the
existence of physical states in midisuperspace models.
The analysis of the wave function of the universe leads
to the conclusion that the classical
curvature singularity present in the evolution of Gowdy models
is removed at the quantum level due to the presence of the
Rarita-Schwinger field.

{\bf File: gowsg2.tex; 30.03.2005}
\end{abstract}


\maketitle

\section{Introduction}
\label{sec1}

Supergravity was originally developed as an elementary
field theory which should avoid the ultraviolet divergencies
and consequently would represent the long awaited
unification of gravity with the remaining fundamental
interactions of nature. Supersymmetry plays an important role in
the development of unification models beyond the standard
model of elementary particles, in the formulation of the
conceptual fundamentals of quantum field theory and quantum
gravity, and more recently in the understanding of
important aspects of superstring theory
 (see, for instance, \cite{reviews1} for reviews of the
 conceptual basis of supersymmetry and supergravity).
Today, supergravity is considered in the first place as an
effective field theory which should describe the low-mass degrees
of freedom of a more fundamental theory, probably the still
unknown M-theory; however, at the moment the only known candidate
for such a theory is superstring theory (see, for example,
\cite{reviews2}).

The classical field equations following from the $N=1$
supergravity Lagrangian were derived in \cite{pi78} by using
the Hamiltonian formalism. There are constraints for each of
the gauge symmetries contained in the theory: spacetime diffeomorphisms,
local Lorentz invariance, and supersymmetry.
One important result that
follows from the analysis of the field equations is that
one of the constraints relates the torsion tensor and the
Rarita-Schwinger field so that it can be used to eliminate
the torsion tensor from the theory.

The canonical quantization
of supergravity is performed in general by applying Dirac's procedure
for constrained systems. One uses the 3+1 decomposition of
the canonical theory to obtain a Hamiltonian
 in which the symmetry generators of the gauge fields are constrained
 by Lagrange multipliers. Then it is postulated that the wave function
is annihilated by all the constraints. In the case of $N=1$ supergravity
there are three constraints: the Hamiltonian   constraint,
the generators of Lorentz rotations and the supersymmetric constraint.
It turns out \cite{tei77} that the Hamiltonian constraint is
identically satisfied once the supersymmetric constraint is fulfilled.
Accordingly, only the Lorentz and supersymmetric constraints are the
central issue.

The study of $N=1$ supergravity models has been limited so far to
minisuperspace models \cite{ho,gc,gc2,pde,opr,maoso,mms98,mac99}
as direct generalizations of
quantum cosmology models. The standard approach to quantum cosmology consists
in canonically quantizing a minisuperspace model which is obtained
by imposing certain symmetry conditions on the metrics allowed on the
spacelike slices of the universe. This procedure, however, reduces the
number of degrees of freedom to a finite number and the problem of
quantization can be attacked by applying the canonical
methods of quantum mechanics
\cite{qc,ryan}. The dynamics of the system is governed by the
Wheeler-DeWitt equation which is a second order differential constraint
equation following from general covariance, and acts
on the wave function of the universe. The most general
minisuperspace models analyzed in the literature correspond to
homogeneous and anisotropic  Bianchi cosmological models. Since
the corresponding metrics depend only on time, the dynamics of
the spacelike 3-dimensional slices becomes trivial, unless an
additional reparametrization is performed. Usually, in the
reparametrization one of the
scale factors of the Bianchi metric is taken as ``internal time"
so that the Wheeler-DeWitt equation generates a wave function of
the universe which explicitly depends on the internal time and the
remaining scale factors. Although this is a quite elegant procedure
which in each case leads to an explicit wave function of the universe,
the main problem regarding the existence of classical initial
singularity remains unsolved. In fact, in all analyzed minisuperspaces
the classical singularity remains at the quantum level.
Moreover, the original hope that the behavior of minisuperspace models
would hold at least qualitatively in the full theory seems to be not realized.
In fact,
in \cite{kucryan} it was shown that even in the simple case
of a microsuperspace (a reduced minisuperspace) contained in the
seed minisuperspace the behavior of the corresponding wave functions
is widely different. The problem of the initial classical singularity has
been attacked alternatively by proposing a wave function for the ground state
\cite{harhaw} or a tunneling effect \cite{vil84}, among other proposals.
In both cases the main idea consists in replacing  the classical
singularity by an {\it ad hoc} postulated universe. The initial
singularity problem remains thus unsolved. The consideration of
an additional Rarita-Schwinger field in the context of supergravity
minisuperspace models (supersymmetric quantum cosmology)
does not solve the problem, and the classical singularity remains.

In all the cases the failure to solve the singularity problem
can be attributed to the fact that, due to the strong symmetry reduction,
only a finite number of degrees of freedom can be considered.
To face this difficulty one
needs to analyze genuine field theories with an infinite number
of degrees of freedom. An option would be to consider milder
symmetry reductions which leave unaffected a specific set of
true local degrees of freedom. These are the so called
midisuperspace models. Such spacetimes have a long history
in general relativity. Indeed, any spacetime which allows
the existence of two commuting Killing vector fields leads
to a real field theory with an infinite number of degrees of
freedom. The field equations in this case can be shown to be
equivalent to the wave equation for a scalar field propagating
in a fictitious flat 2+1-dimensional spacetime \cite{krameretal}.
The local degrees of freedom are contained in the scalar field.

In this work we will consider the specific midisuperspace
described by Gowdy $T^3$ cosmological models \cite{gow71,gow74}
 in the context
of $N=1$ supergravity. We will find an explicit expression
for the wave function of the universe and will show that
the mere consideration of a genuine field theory leads to
a solution of the singularity problem. In fact, the singular
behavior of Gowdy models at a certain time of their evolution
has been investigated in detail at the classical level.
It has been shown that the behavior of the metric
near the singularity
corresponds to the so called ``asymptotically velocity term
dominated" (AVTD) behavior (see, for instance \cite{ber02}, for
a recent review).
We will see that the AVTD singular behavior disappears at the
level of the corresponding wave function of the universe.

This paper is organized as follows.  In Section \ref{sec2} we
revise the canonical formulation of supergravity $N=1$ and analyze
the Lorentz constraint, following closely notations and
conventions of \cite{mamilo}. In Section \ref{sec3} we present the
Gowdy $T^3$ cosmological models and their main properties. Section
\ref{sec4} is devoted to the investigation of the supersymmetric
constraint and the solutions for the wave function of the
universe. Finally, Section \ref{con} contains several final
remarks and the conclusions with indications about different
possibilities of generalizing the results derived in this work.

We adopt the following conventions and notations.
Indices related to world coordinates are denoted by Greek
letters. The ones from the middle of the alphabet,
i. e. $\mu, \ \nu, ...$, run over
0,1,2,3 and the ones from the beginning of the alphabet,
i. e. $\alpha,\ \beta, ...$, represent only spatial
coordinates 1,2,3.
Capital Latin indices, i.e. $A,\ B,...$  can take the values
0,1,2,3 and
the small ones run over 1,2,3; both of them refer to
to components in an orthonormal local frame,
for which we use the local
metric $\eta_{AB}={\rm diag}(+,-,-,-)$.

\section{Canonical formulation of $N=1$ supergravity}
\label{sec2}

The fields of $N=1$ supergravity in 4 dimensions are the vierbein
$e_\mu ^{\ A}$ and the Rarita-Schwinger gravitino $\psi_\mu$,
which is a vector of Majorana spinors.
The corresponding Lagrangian is given by
\begin{equation}
\label{lag1} {\cal{L}}=\frac{1}{2} \sqrt{-g} R- \frac{i}{2}
\epsilon^{\lambda\mu\nu\rho} \bar{\psi}_\lambda \gamma_{5}
\gamma_{\mu} D_{\mu} \psi_{\rho}\, ,
\end{equation}
 where
\begin{equation}
\label{lag2} D_{\nu}=\partial_\nu +\frac{1}{2} \omega_{\nu A B}
\sigma^{AB}\, ,
\end{equation}
is the covariant derivative with the Lorentz generators
$\sigma^{AB}=(1/4)(\gamma^A \gamma^B-\gamma^B \gamma^A)$. For the
$\gamma^A$ matrices we use the real Majorana representation
\be
\label{s4}
\gamma^{0}=\left( \begin{array}{cc} 0&\sigma^2\\
\sigma^2&0\end{array}\right) \, , \,\,
\gamma^{1}=\left( \begin{array}{cc} i\sigma^3&0\\
0&i\sigma^3\end{array}\right)\, , \,\,
\gamma^{2}=\left( \begin{array}{cc} 0&-\sigma^2\\
\sigma^2&0\end{array}\right)\, , \,\,
\gamma^{3}=\left( \begin{array}{cc} -i\sigma^1&0\\
0&-i\sigma^1\end{array}\right) \, ,
\ee in which the anticommutator relation $\{\gamma^A,\gamma^B\} =
2\eta^{AB}$ is satisfied, and $\sigma^i$ are the standard Pauli
matrices. Moreover, $\gamma_5= i\gamma^0\gamma^1\gamma^2\gamma^3$.
In this representation ${\overline \psi}=-i\psi^T \gamma^0$  is known as
the Majorana condition. The (endomorphic) components of the Ricci
rotation coefficients can be obtained from Cartan's first
structure equation ${ d} e^A = - \omega^A_{\ B}\wedge e^B$, and
$\omega_{AB} = \omega_{\mu AB} { d} x^\mu$. Notice that in general
the Ricci rotation coefficients contain a contorsion term.
However, as we mentioned in the introduction the torsion tensor
can be eliminated from the theory by using one of the constraint
equations.

Applying the canonical 3+1 decomposition of spacetime \cite{pi78}
the canonical variables can be chosen to be the spatial components
of the tetrad vectors $e^a_{\ \alpha}$, their conjugate momenta
$p_a^{\ \alpha}$, and the spatial covariant components of the
spinor $\psi_\alpha$. In this case, the corresponding temporal
components turn out to be Lagrange multipliers of the Hamiltonian
which contains only the constraints associated with the three
different types of symmetries of the system:
\begin{equation} H= e^A{}_0\, {\cal H}_A
+ \frac{1}{2} \omega_{0}{}^{AB}\,{\cal J}_{AB}
+ {\overline \psi}_0\, {\cal S} \label{sgham}\, ,
\end{equation}
Here ${\cal H}_A$ contains the usual Hamiltonian and diffeomorphisms
constraints,  ${\cal J}_{AB}$ is the Lorentz constraint and ${\cal S}$
denotes the supersymmetric constraint. According to Dirac's
canonical quantization
procedure for constrained systems, the physical states $\vert\Psi\rangle$
must be annihilated
by the corresponding constraint operators, i.e.
\begin{equation}
{\cal S} \vert\Psi\rangle=0\, , \qquad {\cal H}_A
\vert\Psi\rangle=0\, , \qquad {\cal J}_{AB} \vert\Psi\rangle = 0
\label{phst}\, .
\end{equation}
From the fact that the supergravity operators satisfy Teitelboim's
algebra \cite{tei77} it follows that the condition ${\cal H}_A
\vert\Psi\rangle=0$ is satisfied identically once ${\cal S}
\vert\Psi\rangle=0$ is fulfilled. Consequently, we need to
consider only the Lorentz and supersymmetric constraints.

It is convenient to use,
instead of the gravitino field, the densitized local components
($\psi_0$ is a Lagrange multiplier and so is $\phi_0$ )
\begin{equation}
\phi_a = e\, e_a{}^\alpha \psi_\alpha \label{grav}\, ,
\end{equation}
as the basic fields commuting with all non--spinor variables, where
$e={}^{(3)}e= \det(e_a{}^{\alpha})$.
Moreover, if we choose an $SO(3)$ basis it is possible to show that
all bosonic terms of the Lorentz constraint cancel each other, yielding
\cite{pi78}
\begin{equation}
{\cal J}_{AB} = \frac{1}{2} \phi_{[A}^T \phi_{B]}
\label{lorentz}\, ,
\end{equation}
and the generator of supersymmetry is given by
\begin{equation}
 {\cal S} = \varepsilon^{0\alpha\beta\delta} \gamma_5
\gamma_\alpha D_\beta \psi_\delta \label{susyc}\, ,
\end{equation}
where a factor ordering is usually chosen \cite{mor}.

Let us first analyze the Lorentz condition
${\cal J}_{AB} \vert\Psi\rangle = 0$ that explicitly reads
\begin{equation}
{\cal J}_{AB}\vert\Psi\rangle = \left( \begin{array}{cccc}
 0&0&0&0\\
0&~0&~{\cal J}_{12}&~{\cal J}_{13}\\
0&~-{\cal J}_{12}&~~0&~{\cal J}_{23}\\
0&~-{\cal J}_{13}&~- {\cal J}_{23}&~~0
\end{array}\right)
\left( \begin{array}{c}
 \Psi_I\\
 \Psi_{II}\\
 \Psi_{III}\\
 \Psi_{IV}\end{array}\right)= 0
\label{lor}\, .
\end{equation}
Since this constraint does not affect the component $\Psi_I$, a
particular solution is to keep $\Psi_I$ arbitrary and \be
\Psi_{II}=\Psi_{III}=\Psi_{IV}=0. \label{rfc} \ee This is the
``rest-frame" solution \cite{kaku} in which only the component
$\Psi_I$ remains to be determined by the supersymmetric
constraint. If none of the conditions  (\ref{rfc}) is satisfied,
the Lorentz constraint implies that \cite{Niew81}
\begin{eqnarray}
{\cal J}_{12}\Psi_{III} &=& -{\cal J}_{13}\Psi_{IV}\label{sys1}\, ,\\
{\cal J}_{12}\Psi_{II} &=& {\cal J}_{23}\Psi_{IV}\label{sys2}\, , \\
{\cal J}_{13}\Psi_{II} &=& -{\cal J}_{23}\Psi_{III} \label{sys3}\, .
\end{eqnarray}
Notice that in the representation we are using here the Lorentz
generators ${\cal J}_{ab}$ are $4\times 4$ matrices. Therefore,
the components $\Psi_I, \ \Psi_{II}, \ \Psi_{III}$, and
$\Psi_{IV}$ of the wave function must be considered as $4\times 1$
matrices.

There are two alternative ways to solve the system of algebraic equations given in
(\ref{sys1})--(\ref{sys3}).
The first one consists in taking the components
$\Psi_{II}, \ \Psi_{III}$, and $\Psi_{IV}$ proportional to each other.
The proportionality factors can be absorbed by redefining each component
of the wave function so that this case can be written as
\be
\Psi_{II} = \Psi_{III} = \Psi_{IV} \ .
\label{tcpsi}
\ee
Then from Eqs.(\ref{sys1})-(\ref{sys3}) it follows that
\be
{\cal J}_{12} = - {\cal J}_{13} = {\cal J}_{23} \ ,
\label{tcj}
\ee
a condition which  implies a trivialization of the Lorentz constraint.
Then from the expression (\ref{lorentz}) we can
determine the components of the gravitino field. In the trivial case
(\ref{tcj}) we obtain
\be
\phi_1 = - \phi_2 = \phi_3 \ .
\label{tcphi}
\ee

The second possibility is to solve explicitly the system of equations
(\ref{sys1})--(\ref{sys3}).
The relation between the components of the Lorentz generators can
be solved by representing each component as a product of
$\gamma$-matrices, under the condition that the corresponding
Lorentz algebra is preserved. Alternatively we can use the standard
generators of the ordinary rotation group $O(3)$ as given in \cite{kaku}
\be
{\cal J}_{12}=-i\left( \begin{array}{cccc} 0&0&0&0\\
0&0&1&0\\0&-1&0&0\\0&0&0&0\end{array}\right) \, ,
{\cal J}_{13}=i\left( \begin{array}{cccc} 0&0&0&0\\
0&0&0&-1\\0&0&0&0\\0&1&0&0\end{array}\right)\, , \,\,
{\cal J}_{23}=-i\left( \begin{array}{cccc} 0&0&0&0\\
0&0&0&0\\0&0&0&1\\0&0&-1&0\end{array}\right)\, .
\ee
Solving explicitly the Lorentz constraint (\ref{sys1})--(\ref{sys3}) in
this representation it is easy to see that for each of the vectors
$\Psi_{II}$, $\Psi_{III}$, $\Psi_{IV}$ only one component is different
from zero. That is to say the wave function of the universe can be
represented as
\be
\label{wfu0}
\vert\Psi\rangle=\left( \begin{array}{c}
 \Psi_I\\
 \Psi_{II}\\
 \Psi_{III}\\
 \Psi_{IV}\end{array}\right)
            = E
           \left( \begin{array}{c}
 {\bf a}_0\\
 b_0\\
 c_0\\
 d_0\end{array}\right)\, ,
\ee
where $E$ is a function to be determined, ${\bf a}_0$ is an arbitrary 4--vector and $b_0$, $c_0$, and $d_0$
are arbitrary constants. For the sake of simplicity later on we will consider the special case ${\bf a}_0 =
\{a_0, 0,0,0\}$.  So we have shown that the wave function (\ref{wfu0}) explicitly solves the
Lorentz constraint. The remaining function $E$ and the arbitrary constants $a_0$, $b_0$, $c_0$, and $d_0$
have to be chosen so that the supersymmetric constraint is satisfied.

This ends the analysis of the Lorentz constraint. Notice that in
the ``rest-frame" solution (\ref{rfc}), the wave function of the
universe is a scalar with only one independent component. For the
further investigation of this solution as well as of the
non-trivial solution (\ref{tcpsi}), we need to analyze the
supersymmetric constraint (\ref{susyc}), in which the bosonic part
plays a prominent role. In the next Section we will present the
specific gravitational field which completely determines the
bosonic part of the supersymmetric constraint.

\section{Gowdy $T^3$ cosmological models}
\label{sec3}

Gowdy cosmological models are inhomogeneous time-dependent
solutions of Einstein's vacuum equations with compact Cauchy
spatial hypersurfaces whose topology can be
either $T^3$ or $S^1\times S^2$ \cite{gow71,gow74}. Other particular topologies
are contained in these two as special cases. Here we will focus
on $T^3$ models for which the line element can be written as
\be
ds^2 = e^{-\lambda/2 + \tau/2} (e^{-2\tau} d\tau^2 - d\chi^2)
- e^{-\tau} \left[e^P (d\sigma + Q d\delta)^2 + e^{-P} d\delta^2\right]\ ,
\label{gle}
\ee
where $P$, $Q$, and $\lambda$ depend on the non-ignorable coordinates
$\tau$ and $\chi$. The spatial hypersurfaces $(\tau =$const) are compact
if we require that $0\leq \chi, \sigma, \delta \leq 2\pi$.
The expression in square
brackets depicts the metric on the $T^2$ subspace which is
generated by the commuting Killing vectors $\partial_\sigma$ and
$\partial_\delta$. The coordinate $\chi$ labels the several tori.

When the Killing vectors are hypersurface orthogonal, the general
line element (\ref{gle})
 becomes diagonal with $Q=0$ and the corresponding
cosmological models are called polarized. In this last case,
the subspace $T^2$ corresponds to the spatial surfaces of a
$2+1$ fictitious flat spacetime in which a scalar field,
represented by the metric function $P$, propagates
\cite{ashpierri,ccq1}. The local degrees of freedom contained
in the scalar field are true gravitational degrees of freedom
which cannot be eliminated by a choice of gauge. We are
thus facing a genuine field theory which is a special case
of a midisuperspace model. Notice that the infinite number
of degrees of freedom contained in this midisuperspace model
can be associated with the inhomogeneous character of the
spacetime. If we would neglect the inhomogeneities present
in the model, we would obtain a minisuperspace model with
a finite number of degrees of freedom, probably related to
a Bianchi cosmological model. The general unpolarized case
$(Q\neq 0)$ also corresponds to a midisuperspace model; however,
its interpretation in terms of a dynamical scalar field in
a 2+1 spacetime can not be realized. In this work we will
concentrate on the polarized case $(Q=0)$ where the field equations
can be integrated in general. The general unpolarized case
$(Q\neq 0)$ will be also considered in quite general terms at
the level of the wave function of the universe, although no
exact solution with $Q\neq 0$ will be analyzed due to the
difficulty of the classical field equations.

The vacuum field equations for the general line element (\ref{gle})
can be written as
a set of two second order differential equations for $P$ and
$Q$
\begin{eqnarray}
P_{\tau\tau} - e^{-2\tau} P_{\chi\chi} - e^{2P}(Q_\tau^2 -
e^{-2\tau}Q_\chi^2) &=& 0 \, , \label{t3eqp} \\
Q_{\tau\tau} - e^{-2\tau} Q_{\chi\chi} + 2(P_\tau Q_\tau -
e^{-2\tau}P_\chi Q_\chi) &=& 0 \, ,\label{t3eqq}
\end{eqnarray}
and two first order differential equations for $\lambda$
\begin{eqnarray}
\lambda_\tau &=& P_\tau^2 + e^{-2\tau}P_\chi^2 + e^{2P} (Q_\tau^2
+ e^{-2\tau}Q_\chi^2) \, , \label{t3eqlam1}\\
\lambda_\chi &=& 2(P_\chi P_\tau + e^{2P}Q_\chi Q_\tau)  \, .
\label{t3eqlam2}
\end{eqnarray}
Notice that the function $\lambda$ can be calculated by quadratures once $P$
and $Q$ are known. The system of differential equations (\ref{t3eqp}) and
(\ref{t3eqq}) for $P$ and $Q$ is highly nonlinear, has been investigated
in detail by using numerical methods \cite{varios}, and has been analyzed
analytically only recently in \cite{oqr1,oqr2,nosotros} where several
special solutions have been derived. In the special polarized case,
using the method of separation of variables
it is possible to find the general solution as
\begin{equation} \label{sv10}
Q=0, \ \
P(\tau,\chi)=\sum_{n=0}^{\infty}[A_n\cos{(n\chi)}+B_n\sin{(n\chi)}][C_n
J_0(n e^{-\tau})+ D_n N_0(n e^{-\tau})]\,,
\end{equation}
where $A_n$, $B_n$, $C_n$ and $D_n$ are arbitrary constants and
$J_0$, $N_0$ are Bessel functions. The integration of the function
$\lambda$ is quite cumbersome for the general solution. We will present
in Section \ref{sec4} a particular solution characterized by
a finite number of terms of the sum (\ref{sv10}).

The behavior of Gowdy cosmological models near the singularity is
an important property that has been intensively used to study the
geometric behavior of the initial Big-Bang singularity of our
Universe. In the case of $T^3$ models it can be shown that the
singularity is approached in the limit $\tau \rightarrow \infty$.
It has been proved that all polarized Gowdy models belong to the
class of ``asymptotically velocity term dominated" (AVTD)
solutions and it has been conjectured that the general
(unpolarized) models are also AVTD \cite{mon1}. This conjecture
has been reinforced through the analogy with other midisuperspaces
\cite{her}. The AVTD behavior states that near the singularity
each point in space is characterized by a different spatially
homogeneous cosmology \cite{eardley}. It implies that at the
singularity all spatial derivatives of the field equations can be
neglected and only the temporal behavior is relevant. This leads
to a ``truncated" set of differential equations which in the case
of $T^3$ models can be obtained from Eqs.(\ref{t3eqp}) -
(\ref{t3eqlam2}) by neglecting all the derivatives with respect to
the spatial coordinate $\chi$. It is easy to see that the general
solution to this ``truncated" system is given by
\cite{Berger,nosotros} \be P_{AVTD} = \ln [ a( e^{-c\tau} + b^2
e^{c\tau})]\ , \quad Q_{AVTD} = \frac{b}{a(e^{-2c\tau} + b^2)} + d
\ , \quad \lambda_{AVTD} = \lambda_0 + c^2 \tau \ , \label{avtd}
\ee where $a ,\ b, \ c$, $d$, and $\lambda_0$ can be considered as
arbitrary real constants. The singularity situated at
$\tau\rightarrow \infty$ is characterized by a blow up of the
curvature which is determined by the behavior of the AVTD solution
(\ref{avtd}).

For the calculation of the supersymmetric constraint we need
explicitly the connection in an orthonormal basis. The
structure of the line element (\ref{gle}) suggests the
following choice for the vierbein
\begin{equation}
\label{t6} e^0 = e^{-(\lambda+3\tau)/4} d \tau \, , \quad e^1 =
e^{-(\lambda-\tau)/4} d \chi\, , \quad e^2 = e^{(P-\tau)/2}
(d\sigma + Q d\delta) \, , \quad e^3 = e^{-(P+\tau)/2} d\delta \,
,
\end{equation}
which satisfies the orthonormality condition
$g^{\mu\nu}e^{A}{}_{\mu}e^{B}{}_{\nu}=\eta^{AB}$ with $e^A =
e^A_{\ \mu} d x^\mu$. In this tetrad the non-vanishing
components of the connection $\omega_{\mu AB}$ are
\begin{eqnarray}
\label{ome} \omega_{0\hat{0}\hat{1}}&=&-\frac{1}{4}\lambda_{\chi}
\, , \qquad \qquad \qquad~
\omega_{0\hat{2}\hat{3}}=-\frac{1}{2}e^{\tau+P}Q_{\tau}\, ,
\nonumber\\
\omega_{1 \hat{0}\hat{1}}&=&-
\frac{1}{4}e^{\tau}(1-\lambda_{\tau})\ , \quad \qquad
\omega_{1\hat{2}\hat{3}}=\frac{1}{2}e^{P}Q_{\chi}\ ,\nonumber\\
\omega_{2\hat{0}\hat{2}}&=&\frac{1}{2}e^{\frac{\tau}{4}+\frac{\lambda}{4}+\frac{P}{2}}(1-P_{\tau})\
, ~\quad
\omega_{2\hat{3}\hat{0}}=-\frac{1}{2}e^{\frac{\tau}{4}+\frac{\lambda}{4}}Q_{\tau}\ , \nonumber\\
\omega_{2 \hat{1}\hat{2}}&=&
\frac{1}{2}e^{-\frac{3}{4}\tau+\frac{\lambda}{4}+\frac{P}{2}}P_{\chi}
\ ,\qquad ~~~
\omega_{2\hat{1}\hat{3}}=\frac{1}{2}e^{-\frac{3}{4}\tau+\frac{\lambda}{4}+\frac{3}{2}P}Q_{\chi}\ , \\
\omega_{3\hat{0}\hat{2}}&=&\frac{1}{2}e^{\frac{\tau}{4}+\frac{\lambda}{4}+P}[(1-e^{-\tau}P_{\tau})Q-Q_{\tau}]
\ , \nonumber\\
\omega_{3 \hat{0}\hat{3}}&=&-\frac{1}{2}e^{\frac{\tau}{4}+\frac{\lambda}{4}}[e^{2P}Q_{\tau}Q-(P_{\tau}+1)]\ , \nonumber\\
\omega_{3\hat{1}\hat{2}}&=&\frac{1}{2}e^{-\frac{3}{4}\tau+\frac{\lambda}{4}+\frac{P}{2}}(P_{\chi}Q+Q_{\chi})
\ , \nonumber\\
\omega_{3\hat{1}\hat{3}}&=&-\frac{1}{2}e^{-\frac{3}{4}\tau+\frac{\lambda}{4}-\frac{P}{2}}(P_{\chi}-e^{2P}Q_{\chi}Q)
\ .
\nonumber
\end{eqnarray}
where the hat refers to indices associated to the local
orthonormal basis.

\section{Physical states}
\label{sec4}

In this section we analyze the remaining supersymmetric constraint.
The densitized local components of the gravitino field (\ref{grav})
depend on the components of the local Gowdy tetrad (\ref{t6}). Noting
that in this case $e={}^{(3)}e= \det(e_a{}^{\alpha})= \exp[(\lambda+3\tau)/4]$,
we obtain from Eq.(\ref{grav}) that
\be
\psi_1 = e^{-\frac{1}{2}(\lambda+\tau)} \phi_1 \ , \quad
\psi_2 = e^{-\frac{1}{4}(\lambda+5\tau)} \phi_2 \ , \quad
\psi_3 = e^{-\frac{1}{4}(\lambda+5\tau + 2P)}(\phi_3 + e^P Q\phi_2) \ .
\label{psis}
\ee
With these values for the gravitino field and the
connection components (\ref{ome}), it is now straightforward to
determine the explicit form of the supersymmetric constraint
(\ref{susyc}) which after lengthly calculations can be written
as
\be
{\cal S} = e^{-\frac{1}{4}(\lambda+7\tau)} {\cal S}_1 +
e^{-\frac{1}{2}(\lambda+2\tau-P)} {\cal S}_2 \ ,
\label{susyc1}
\ee
\be
{\cal S}_1 = i\, \gamma^0\gamma^1\left[
\gamma^2\phi_2\left(\partial_\chi -\frac{1}{4}\lambda_\chi +\frac{1}{4} P_\chi\right)
+\gamma^3\phi_3\left(\partial_\chi -\frac{1}{4}\lambda_\chi -\frac{1}{4} P_\chi\right)
+ \frac{1}{2} \gamma^3\phi_2 e^P Q_\chi\right] \ ,
\label{s1}
\ee
\be
\label{s2}
{\cal S}_2 = i\,  \gamma^0 \{[e^{-P}(\gamma^2\gamma^3\phi_3 -\gamma^1\gamma^2\phi_1)+
Q(\gamma^1\gamma^3\phi_1+\gamma^2\gamma^3\phi_2)  ]\partial_\sigma
-(\gamma^2\gamma^3\phi_2 +\gamma^1\gamma^3\phi_1)\partial_\delta\} \ .
\ee
The component  ${\cal S}_1$ contains all the terms which include
dependence on the non-ignorable spatial coordinate $\chi$. The second
components contains derivatives with respect to the spatial coordinates
$\sigma$ and $\delta$. Since the classical spacetime metric depends only
on the spatial coordinate $\chi$, one could expect a similar dependence
for the wave function  $\vert\Psi\rangle$. In such a case, the action of
${\cal S}_2$ on the wave function would vanish and we would need to consider
only the first term ${\cal S}_1$. However, we will analyze here
the most general case allowed by the supersymmetric constraint
${\cal S}\vert\Psi\rangle =0$.

According to the discussion of Section \ref{sec2}, there are two
different types of solutions to the Lorentz constraint: the
rest-frame (\ref{rfc}), and the trivial solution
(\ref{tcpsi}). The next step is to solve the supersymmetric
constraint for each of these special solutions. However, as
mentioned above the rest-frame solution leads to scalar wave
functions. This implies that the wave function contains only
bosonic degrees of freedom and, according to \cite{cfop}, the
corresponding states cannot be physical. In fact, physical states
must contain fermionic degrees of freedom and have infinite
 number of modes. In the case of the trivial solution (\ref{tcpsi}),
its non-physical character could have been expected from the
fact that the trivialization of the Lorentz constraint as given
in Eq.(\ref{tcj}) does not fulfill Teitelboim's algebra \cite{tei77}.
  Thus, we are left only the non-trivial solution
(\ref{wfu0}). On the one hand, this special non-trivial choice of
the gravitino field guarantees that fermionic degrees of freedom
will enter the final form of the wave function and, on the other
hand, it satisfies identically the Lorentz constraint in the sense
that it leads to an expression for the wave function which
involves the fermionic variable in a manifestly Lorentz invariant
combination.

From the general expression for the supersymmetric operator
(\ref{susyc1}), (\ref{s1}), and (\ref{s2}), we obtain
\begin{eqnarray} {\cal S}_1 &=& \left(\partial_\chi
-\frac{1}{4}\lambda_\chi +\frac{1}{4} P_\chi\right)\Gamma^1
     +\left(\partial_\chi -\frac{1}{4}\lambda_\chi -\frac{1}{4} P_\chi\right)\Gamma^2
+ \frac{1}{2}  e^P Q_\chi  \Gamma^3\ , \label{s1nt} \\
 {\cal S}_2 &=& ( e^{-P}\Gamma^4 +Q \Gamma^5 )
\partial_\sigma -\Gamma^5\partial_\delta\ \label{s2nt}\, ,
\end{eqnarray}
where, in order to solve them, we use the following
realization for the product of the $\gamma$ matrices with
\begin{equation}
\phi_1 = -i\gamma^3 \, , \qquad \phi_2 =-i \gamma^1 \, , \qquad
\phi_3 = -i \gamma^0 \label{papi} \, .
\end{equation}
in terms of the following $\Gamma^i$ matrices \be
  \Gamma^1= \left( \begin{array}{cccc}
 1&0&0&0\\
0&1&0&0\\
0&0&-1&0\\
0&0&0&-1
\end{array}\right)
\ , \quad
 \Gamma^2= \left( \begin{array}{cccc}
 0&1&0&0\\
-1&0&0&0\\
0&0&0&1\\
0&0&-1&0
\end{array}\right)
\ , \quad
 \Gamma^3= \left( \begin{array}{cccc}
 0&0&-1&0\\
0&0&0&1\\
-1&0&0&0\\
0&1&0&0
\end{array}\right)\ ,
\ee
\be
  \Gamma^4= \left( \begin{array}{cccc}
 0&1&1&0\\
-1&0&0&-1\\
-1&0&0&-1\\
0&1&1&0
\end{array}\right)
\ , \quad
 \Gamma^5= \left( \begin{array}{cccc}
 0&-1&0&1\\
1&0&1&0\\
0&1&0&1\\
1&0&-1&0
\end{array}\right)
\ .
\ee

The physical states are given as the non-trivial solutions of the equation ${\cal S} \vert\Psi\rangle = 0$.
The investigation of this equation in minisuperspace models \cite{mamilo} showed
that there are no non-trivial solutions and, therefore, sometimes it is believed that supergravity
$N=1$ is an uninteresting theory with no physical states.
A less radical conclusion would be
that minisuperspace models are unphysical due to the fact that the
strong symmetry reduction, which
leads to a system with finite number of degrees of freedom,
does not allow the existence of non-trivial
physical solutions to the supersymmetric constraint.
In the present work we are dealing with a midisuperspace with an infinite number of degrees of freedom
and therefore the last explanation does not hold. Indeed, in this case the non existence of genuine
physical states would indicate a very serious difficulty for $N=1$ supergravity. We will show that
this is not the case.

From the above considerations it is clear that a family of
physical states can be obtained from wave functions which
simultaneously satisfy the constraints ${\cal S}_1
\vert\Psi\rangle = 0$ and ${\cal S}_2 \vert\Psi\rangle = 0$. As
mentioned above, if in accordance to the functional dependence of the
classical metric we suppose that the wave function depends only on
the spatial coordinate $\chi$, the constraint ${\cal S}_2
\vert\Psi\rangle = 0$ is identically satisfied and we only need to
solve the set of differential equations following from ${\cal S}_1
\vert\Psi\rangle = 0$. If we limit ourselves to wave functions
which are independent of the coordinates $\sigma$ and $\delta$, we
guarantee that ``anomalies'' do not appear. Indeed, the classical
symmetries associated with the Killing vectors
$(\partial_\sigma)^\mu$ and $(\partial_\delta)^\mu$ does not hold
at the quantum level if we find wave functions which depend on
these coordinates. This would be an indication of the existence of
anomalies at the quantum level. We will show in the next
subsections that it is not necessary to assume independency of
$\sigma$ and $\delta$. We will see that the differential equations
following from the constraint ${\cal S}\vert\Psi\rangle=0$ can be
solved by applying the method of separation of variables and that
the resulting compatibility conditions for the wave function of
the universe eliminate the possibility of existence of anomalies.

\subsection{The polarized case}
\label{subsec1}

As mentioned in Section \ref{sec3}, the polarized case of Gowdy
models corresponds to the limit $Q=0$ of the line element
(\ref{gle}), and the differential equation for the function $P$
allows the general solution given in Eq.(\ref{sv10}). The
supersymmetric constraint ${\cal S} \vert\Psi\rangle = 0$ leads to the
following set of first order partial differential equations
\begin{eqnarray}
\label{gc1}
 & &\partial_\chi (\Psi_I + \Psi_{II})
-\frac{1}{4}(\lambda_{\chi}-P_{\chi})\Psi_I
-\frac{1}{4}(\lambda_{\chi}+P_{\chi})\Psi_{II}\nonumber \\
& &+ e^{-\frac{1}{4}(\lambda -2P-3\tau)} \left[
e^{-P}\partial_\sigma(\Psi_{II}+\Psi_{III}) +
\partial_\delta (\Psi_{II} -\Psi_{IV}) \right] =0 \,,
\end{eqnarray}
\begin{eqnarray}
\label{gc2}
& &\partial_\chi (\Psi_{II} - \Psi_{I})
-\frac{1}{4}(\lambda_{\chi}-P_{\chi})\Psi_{II}
+\frac{1}{4}(\lambda_{\chi}+P_{\chi})\Psi_{I}\nonumber \\
& & - e^{-\frac{1}{4}(\lambda -2P-3\tau)} \left[
e^{-P}\partial_\sigma(\Psi_{I}+\Psi_{IV})
+\partial_\delta (\Psi_{I} +\Psi_{III}) \right] =0 \,,
\end{eqnarray}
\begin{eqnarray}
\label{gc3}
& & \partial_\chi (\Psi_{IV} - \Psi_{III})
+\frac{1}{4}(\lambda_{\chi}-P_{\chi})\Psi_{III}
-\frac{1}{4}(\lambda_{\chi}+P_{\chi})\Psi_{IV}\nonumber \\
& &- e^{-\frac{1}{4}(\lambda -2P-3\tau)} \left[
e^{-P}\partial_\sigma(\Psi_{I}+\Psi_{IV})
+\partial_\delta (\Psi_{II} +\Psi_{IV}) \right] =0 \,,
\end{eqnarray}
\begin{eqnarray}
\label{gc4}
& & -\partial_\chi (\Psi_{III} + \Psi_{IV})
+\frac{1}{4}(\lambda_{\chi}-P_{\chi})\Psi_{IV}
+\frac{1}{4}(\lambda_{\chi}+P_{\chi})\Psi_{III}\nonumber \\
& &+ e^{-\frac{1}{4}(\lambda -2P-3\tau)} \left[
e^{-P}\partial_\sigma(\Psi_{II}+\Psi_{III}) +
\partial_\delta (-\Psi_{I} +\Psi_{III}) \right] =0 \,.
\end{eqnarray}
First, we consider the special case where the components of the
wave function are independent of the spatial coordinates $\sigma$ and $\delta$.
This means that the constraint ${\cal S}_2\vert\Psi\rangle=0$ is identically satisfied
and the last set of equations reduce to a system of ordinary differential
equations. To find solutions to the resulting
system it is natural to use the exponential functions
as an  ansatz for each of the components of the wave function. It is then
straightforward to show that the general solution is given by
\begin{equation}
\vert\Psi\rangle=\left(\begin{array}{c}
\Psi_I\\
\Psi_{II}\\
\Psi_{III}\\
\Psi_{IV}
\end{array}\right)= e^{\frac{1}{4}(\lambda\mp iP)}\left(\begin{array}{c}
a_0 \\
b_0 \\
c_0 \\
d_0\end{array}\right) \label{wfu1}\, ,
\end{equation}
where $a_0,\ b_0,\ c_0,$ and $d_0$ are arbitrary constants satisfying
the relationships
\be
a_0^2 + b_0^2 =0\ , \qquad c_0^2 + d_0^2 =0\ ,
\label{conds1}
\ee
for which we choose the solution $a_0=\pm i b_0$ and $c_0=\mp i d_0$.
This wave function of the universe represents a {\it physical state}
 for the $N=1$ supergravity Gowdy model. It is interesting to notice
 that in this particular case we were able to completely integrate
 the system of differential equations and all the components of
 $\vert\Psi\rangle$ are explicitly given in terms of the ``classical''
 functions $P$ and $\lambda$. Moreover, the metric function $P$ enters
 the wave function of the universe in a way very similar to a phase
 which does not affect the physical significance of the solution.
 Indeed, if we define the ``absolute value'' of the wave function of
 the universe  as $\vert\Psi\rangle \vert\Psi\rangle^* =
 \Psi_I\Psi_I^* +  \Psi_{II}\Psi_{II}^* +  \Psi_{III}\Psi_{III}^*
  +  \Psi_{IV}\Psi_{IV}^*$, we obtain from Eq.(\ref{wfu1})
\be
\vert\Psi\rangle \vert\Psi\rangle^* =
(a_0a_0^*+b_0b_0^*+c_0c_0^*+d_0d_0^*)e^{\frac{1}{2}\lambda} \ .
\ee
Notice that this ``absolute value'' can not be associated with a
probability density (see the discussion in Section \ref{con}).
The behavior of the wave function of the universe is thus entirely
dictated by the metric function $\lambda$. Let us recall that
the function $P$ obeys a linear differential equation and also
determines the behavior of $\lambda$ through Eqs.(\ref{t3eqlam1})
and (\ref{t3eqlam2}) which are nonlinear. We see that
the nonlinear sector of the classical field equations enters
the final form of the wave function of the universe, whereas the
linear sector appears as a phase only.

\begin{figure}[t]
\begin{center}
\includegraphics*[scale=0.5]{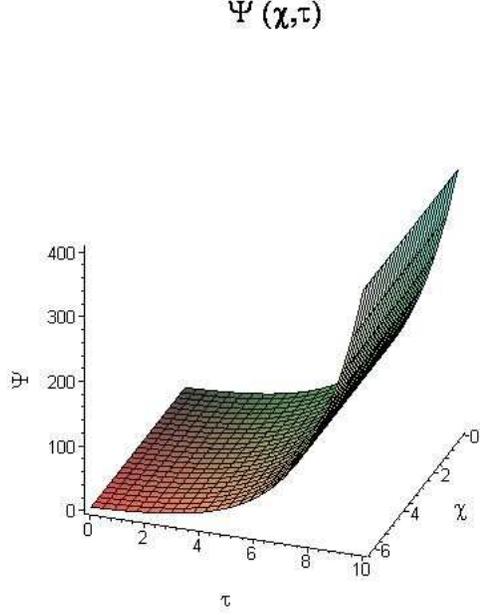}
\end{center}
\caption{ The wave function of the universe is formally defined
for the AVTD solution as $\vert\Psi\rangle_{AVTD} \sim
\exp(\lambda_{AVTD}/2) = \exp[(\lambda_0+c^2\tau)/2]$, according
to Eq.(\ref{avtd}). The graphic shows the singular behavior of
$\vert\Psi\rangle_{AVTD}$, for $\tau\rightarrow\infty$, with
$c=1$, and $\lambda_0=1$.\label{figure1}}
\end{figure}

With the wave function of the universe we can analyze the problem
of the cosmological singularity. Recall that the classical
spacetime is characterized by a singularity at
$\tau\rightarrow\infty$ where the metric functions behave
according to the AVTD solution (\ref{avtd}). In particular, the
metric function $\lambda_{AVTD}$ diverges linearly as
$\tau\rightarrow\infty$. This singular behavior is illustrated in
Fig.~\ref{figure1}. Let us consider a special solution contained
in (\ref{sv10}) with $D_n=0$ and $C_n=1$. We choose this special
case for the sake of simplicity and because it reproduces the
structure of the general solution. Consider also the case with
$n=1,2$, i.e. \be P=A_1 \cos{\chi}\,J_0(e^{-\tau})+ A_2
\cos{2\chi}\,J_0(2e^{-\tau})\,. \label{p12} \ee Notice that the
first term $n=0$ leads to a constant value of $P$ which can be
absorbed in the metric through a coordinate transformation and
leads to the Minkowski metric. For this reason we do not consider
this term in the series (\ref{sv10}). Using the expression
(\ref{p12}) the field equations (\ref{t3eqlam1}) and
(\ref{t3eqlam2}) can be integrated and yield
\begin{eqnarray}
\lambda&=&-A_1^2e^{-\tau}J_0(e^{-\tau})J_1(e^{-\tau})\sin^2{\chi}+
\frac{8}{3}A_1A_2e^{-\tau}J_0(2e^{-\tau})J_1(e^{-\tau})\cos^3{\chi}+\nonumber\\&+&
\frac{4}{3}A_1A_2e^{-\tau}J_0(e^{-\tau})J_1(2e^{-\tau})\cos{\chi}(\cos{2\chi}-2)-
2A_2^2e^{-\tau}J_0(2e^{-\tau})J_1(2e^{-\tau})\sin^2{2\chi}-\nonumber\\&-&
\frac{1}{2}A_1^2e^{-2\tau}[J_1(e^{-\tau})^2-J_0(e^{-\tau})J_2(e^{-\tau})]-
2A_2^2e^{-2\tau}[J_1(2e^{-\tau})^2-J_0(2e^{-\tau})J_2(2e^{-\tau})]\,.
\label{lam12}
\end{eqnarray}

\begin{figure}[t]
\begin{center}
\includegraphics*[scale=0.5]{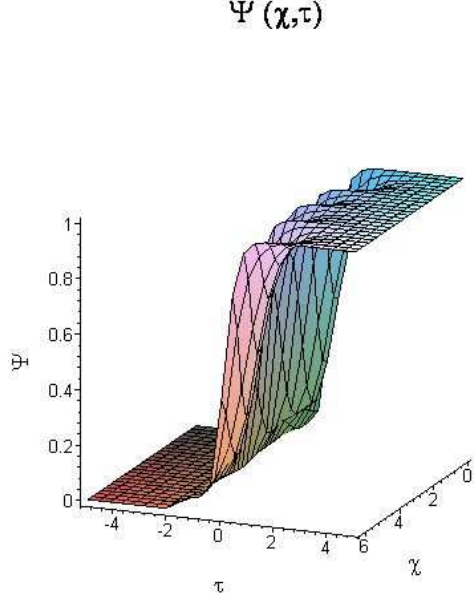}
\end{center}
\caption{ Behavior of the wave function of the universe for the
special solution (\ref{p12}), and (\ref{lam12}). The AVTD singular
behavior for $\tau\rightarrow\infty$, has disappeared and it is
replaced by a well-behaved function with a constant
value.\label{figure2}}
\end{figure}

The wave function of the universe for this special case is
depicted in Fig.~\ref{figure2}, where we can see that it is
regular for all values of $\chi$ and $\tau$. As the classical
singularity ($\tau\rightarrow\infty$) is approached, the wave
function of the universe remains constant and finite. This result
shows that the classical cosmological singularity of this
particular Gowdy model has been removed after its canonical
quantization in the context of $N=1$ supergravity. The presence of
the fermionic field is thus sufficient to solve the singularity
problem at the quantum level in  this particular midisuperspace
model. To obtain this result we have considered in (\ref{p12})
only the first two nontrivial terms of an infinite series. If we
consider further terms, the expression for the corresponding
function $\lambda$ become rather cumbersome and difficult to
handle. Nevertheless, since all higher terms contain only Bessel
functions, the mathematical structure of the  function $\lambda$
resembles that of Eq.(\ref{lam12}) so that for further terms we
can expect again wave functions of the universe which are free of
singularities. To confirm in an invariant way that no
singularities appear in the wave function of the universe we have
analyzed the curvature spatial scalar $R = e^{\ a}_\alpha
R^\alpha_{\ a}$, which characterizes the structure of the constant
time hypersurface where the wave function of the universe is
defined. In the present case we find that
$R=-(1/2)\exp[(\lambda-\tau)/2]P_\chi^2$, and a numerical analysis
of this scalar shows that it is regular everywhere.

Let us now consider the general case in which the components of
the wave function depend also on the spatial coordinates
$\sigma$ and $\delta$. Using the method of separation of
variables, it is easy to show that the general solution to the system
(\ref{gc1})-(\ref{gc4}) must have the form
\begin{equation}
\vert\Psi\rangle=\left( \begin{array}{c}
 \Psi_I\\
 \Psi_{II}\\
 \Psi_{III}\\
 \Psi_{IV}\end{array}\right)
            = e^{ m\sigma + n\delta + g(\chi) }
           \left( \begin{array}{c}
 a_0\\
 b_0\\
 c_0\\
 d_0\end{array}\right)
\label{wfu2}\, ,
\end{equation}
where $m$ and $n$ are constants which arise from the separation of
variables, and $g(\chi)$ is a function to be determined.
Furthermore, $a_0,\ b_0,\ c_0,$ and $d_0$ are arbitrary constants.
Inserting this ansatz for the wave function into Eqs.(\ref{gc1})
and (\ref{gc2}), we obtain one single differential equation
\be
g_\chi=\frac{1}{4}\lambda_{\chi}\mp \frac{i}{4}P_{\chi}+
e^{-\frac{1}{4}(\lambda-2P-3\tau)}\left[e^{-P}m\frac{a_{0}+
d_{0}}{b_{0}-a_{0}}+n\frac{a_{0}+ c_{0}}{b_{0}-a_{0}}\right]\ ,
\label{eqg1}
\ee
if the following relationships are satisfied
\be
c_0(b_0-a_0) + d_0 (a_0+b_0) =0\ , \quad d_0(a_0-b_0) +
c_0(a_0+b_0) =0\ . \label{conds2}
\ee
Similarly, from
Eqs.(\ref{gc3}) and (\ref{gc4}) we obtain
\be
g_\chi=\frac{1}{4}\lambda_{\chi}\mp \frac{i}{4}P_{\chi}+
e^{-\frac{1}{4}(\lambda-2P-3\tau)}
\left[e^{-P}m\frac{b_{0}+c_{0}}{c_{0}+d_{0}}-n\frac{a_{0}-c_{0}}{c_{0}+d_{0}}\right]
\ , \label{eqg2}
\ee
when the conditions (\ref{conds2}) are
fulfilled. Notice that Eqs.(\ref{conds2}) imply that $a_0^2 +
b_0^2 =0$ and $c_0^2 + d_0^2 =0$ so that we can choose \be a_0 =
\pm i b_0\ , \qquad c_0 = \mp i d_0 \ , \label{sol1} \ee as the
general solution for the system (\ref{conds2}). Comparing term by
term the two equations for the function $g_\chi$ given above we
find two  new conditions on the constants $b_0$ and $d_0$, i.e.
\be
(a_0+d_0)(c_0+d_0)=(b_0+c_0)(b_0-a_0)\ , \quad
(a_0+c_0) (c_0+d_0) = (a_0-c_0)(a_0-b_0) \ .
 \label{conds3}
\ee
Notice that these conditions follow from the terms containing
the constants ``$m$'' and ``$n$'' in Eqs.(\ref{eqg1}) and (\ref{eqg2})
so that they are valid only if $m\neq 0$ and $n\neq 0$.
It is now easy to see that Eqs.(\ref{conds3}) does not allow any solutions
compatible with (\ref{sol1}). Consequently, the only compatible solution
is  for $m=0=n$, a result which implies that an explicit dependence on
the spatial coordinates $\sigma$ and $\delta$ is not allowed and, therefore,
the existence of anomalies in the wave function of the universe is
completely excluded.

Under the above considerations, the resulting differential equation for $g(\chi)$
can easily be integrated and yields $g(\chi) = (\lambda \mp i P)/4$. The imaginary
part can again be considered as a phase and, therefore, the behavior of the wave
function of the universe is completely dictated by the behavior of
the metric function $\lambda$ only. The explicit final expression for
$\vert\Psi\rangle$ coincides with the one of the former case given in
Eq.(\ref{wfu1}).

\subsection{The unpolarized case}
\label{subsec2}

In this Section we will analyze the general unpolarized $T^3$ Gowdy
model $(Q\neq 0)$.
No general solution to the classical field equations is known
in this case because of their highly nonlinear character. In fact,
only a few exact highly nontrivial solutions are known in the literature
\cite{oqr1,oqr2,nosotros,procmike}. Let an example be given for the functions
$P$, $Q$, and $\lambda$ satisfying the field equations (\ref{t3eqp})--
(\ref{t3eqlam2}). This is all the information we need to know in order
to investigate the supersymmetric constraint $S\vert\Psi\rangle=0$
which yields the following set of equations
\begin{eqnarray}
\label{gcu1}
 & &\partial_\chi (\Psi_I + \Psi_{II})
-\frac{1}{4}(\lambda_{\chi}-P_{\chi})\Psi_I
-\frac{1}{4}(\lambda_{\chi}+P_{\chi})\Psi_{II}
-\frac{1}{2}e^P Q_\chi \Psi_{III}
\nonumber \\
& &+ e^{-\frac{1}{4}(\lambda -2P-3\tau)} \left[
e^{-P}\partial_\sigma(\Psi_{II}+\Psi_{III})
- Q\partial_\sigma (\Psi_{II} - \Psi_{IV})
+\partial_\delta (\Psi_{II} -\Psi_{IV}) \right] =0 \,,
\end{eqnarray}
\begin{eqnarray}
\label{gcu2}
& &\partial_\chi (\Psi_{II} - \Psi_{I})
-\frac{1}{4}(\lambda_{\chi}-P_{\chi})\Psi_{II}
+\frac{1}{4}(\lambda_{\chi}+P_{\chi})\Psi_{I}
+\frac{1}{2}e^P Q_\chi \Psi_{IV} \nonumber \\
& & - e^{-\frac{1}{4}(\lambda -2P-3\tau)} \left[
e^{-P}\partial_\sigma(\Psi_{I}+\Psi_{IV})
-Q\partial_\sigma (\Psi_{I} +\Psi_{III})
+\partial_\delta (\Psi_{I} +\Psi_{III}) \right] =0 \,,
\end{eqnarray}
\begin{eqnarray}
\label{gcu3}
& & \partial_\chi (\Psi_{IV} - \Psi_{III})
+\frac{1}{4}(\lambda_{\chi}-P_{\chi})\Psi_{III}
-\frac{1}{4}(\lambda_{\chi}+P_{\chi})\Psi_{IV}
-\frac{1}{2}e^P Q_\chi \Psi_{I} \nonumber \\
& &- e^{-\frac{1}{4}(\lambda -2P-3\tau)} \left[
e^{-P}\partial_\sigma(\Psi_{I}+\Psi_{IV})
-Q\partial_\sigma (\Psi_{II} +\Psi_{IV})
+\partial_\delta (\Psi_{II} +\Psi_{IV}) \right] =0 \,,
\end{eqnarray}
\begin{eqnarray}
\label{gcu4}
& & -\partial_\chi (\Psi_{III} + \Psi_{IV})
+\frac{1}{4}(\lambda_{\chi}-P_{\chi})\Psi_{IV}
+\frac{1}{4}(\lambda_{\chi}+P_{\chi})\Psi_{III}
+\frac{1}{2}e^P Q_\chi \Psi_{II} \nonumber \\
& &+ e^{-\frac{1}{4}(\lambda -2P-3\tau)} \left[
e^{-P}\partial_\sigma(\Psi_{II}+\Psi_{III})
+ Q \partial_\sigma (\Psi_{I} -\Psi_{III})
- \partial_\delta (\Psi_{I} -\Psi_{III}) \right] =0 \,.
\end{eqnarray}
It is clear that an exponential
ansatz with separation of variables similar to the one used
in the previous case in Eq.(\ref{wfu2}) will lead to a system of
ordinary differential equations and a set of algebraic
equations. The analysis of the resulting equations is similar
to the one performed in the last subsection for the polarized
case. In a similar manner, it is possible to show that no
anomalies are allowed in the wave function of the universe
which therefore turns out to depend on the spatial coordinate
$\chi$ only.
The resulting wave function of the universe
can be expressed as
\begin{equation}
\vert\Psi\rangle=\left( \begin{array}{c}
 \Psi_I\\
 \Psi_{II}\\
 \Psi_{III}\\
 \Psi_{IV}\end{array}\right)
            = e^{ h(\chi) }
            \left( \begin{array}{c}
 a_0\\
 b_0\\
c_0\\
 d_0 \end{array}\right)
\label{wfu3}\, ,
\end{equation}
where the set of constants must satisfy the conditions \be
c_0(b_0-a_0) + d_0 (a_0+b_0) =0\ , \quad d_0(a_0-b_0) +
c_0(a_0+b_0) =0\ , \label{conds4} \ee which allow the solution
$a_0 = \pm i b_0$, $c_0 = \mp i d_0$. For the sake of simplicity
we consider the case in which $b_0$ and $d_0$ are real constants.
Then, the function $h(\chi)$ can be put in the form

\be h(\chi) = \frac{1}{4}\lambda -\frac{d_0}{4b_0} \int e^P
Q_\chi\, d\chi \mp \frac{i}{4}\left( P + \frac{d_0}{b_0} \int e^P
Q_\chi\, d\chi \right) \ .  \label{h} \ee

This function together with the conditions (\ref{conds4})
represent an explicit solution of the supersymmetric constraint
and show that in the unpolarized case there exist nontrivial
physical states. The imaginary part of $h(\chi)$ can again be
interpreted as a phase which does not affect the behavior of the
wave function of the universe (\ref{wfu3}). The behavior of the
real part of $h(\chi)$ is governed by the behavior of the metric
function $\lambda$ and an integral which involves the functions
$P$ and $Q$. Hence, we need the explicit form of the metric in
order to analyze the wave function of the universe. As mentioned
before, it is very difficult to solve the system of differential
equations which follow from Einstein's vacuum field equations. To
analyze the behavior of the wave function near the singularity it
would be necessary to apply numerical methods for handling the
metric functions. This is a task which is outside the scope of the
present work.

\section{Final remarks and conclusions}
\label{con}

Our analysis of the supersymmetric constraint presented in
the last sections relies on a very specific foliation of
spacetime. In fact, the general form of the constraint
${\cal S} = \varepsilon^{0\alpha\beta\delta} \gamma_5
\gamma_\alpha D_\beta \psi_\delta$ fixes at the
very beginning  the
time coordinate $x^0$ so that it does not appear as
a dynamical variable in the further analysis.
Here we have chosen $x^0=\tau$ which is the natural
time coordinate in Gowdy cosmological models.
The canonical quantization formalism forces us
to ``freeze'' this time during the entire analysis
and it reappears explicitly only in the solutions for
the wave function of the universe, where we interpret
it as a label which associates a particular value
of the wave function to a  different spatial slice.
It is in this sense that we can say that we
know explicitly the wave function of the universe
at each moment of time. And it is in this context that
we were  able
to investigate the behavior of the wave function
near the cosmological singularity.

Consequently, the problem of the ``frozen'' time
in our analysis cannot be solved in the context
of the canonical quantization formalism. The
common concern about the use of this formalism
is whether the final result of the quantization
(in our case, the wave function of the universe)
depends on the choice of a particular foliation. To clarify this
question in the present case let us consider a different
quite general foliation. Consider a new time coordinate $t$
defined by
\be
dt = e^{-\frac{1}{4}(\lambda + 3\tau)} d\tau \ .
\label{ttran}
\ee
Then, from the general line element (\ref{gle}) we obtain
\be
ds^2 =  d t^2 - e^{2\Lambda} d\chi^2 -
 e^F \left(e^P d\sigma ^2 + e^{-P} d\delta^2\right)\ ,
\label{gle1} \ee where $\Lambda=\Lambda(t,\chi)$ and
$F=F(t,\chi)$. For the sake of simplicity we are considering here
the polarized case only $(Q=0)$. In this particular
parametrization the lapse function becomes a constant. We apply
now the canonical quantization procedure of supergravity with a
foliation determined by $x^0=t=$const. Clearly, this is a quite
radical change of foliation when compared with the original one
$(\tau=$const). To calculate the supersymmetric constraint we
proceed as in Section \ref{sec4} for the non-trivial solution
(\ref{wfu0}). Then we obtain \be {\cal S} = e^{\Lambda+2F} {\cal
S}_1 + e^{ \frac{1}{2}(4\Lambda+3F+P)} {\cal S}_2 \ ,
\label{susycnt} \ee with \begin{eqnarray} {\cal S}_1 &=&
\left(\partial_\chi + \Lambda_\chi +\frac{1}{4} P_\chi
+\frac{7}{4}F_\chi \right)\Gamma^1 +\left(\partial_\chi
+\Lambda_\chi -\frac{1}{4} P_\chi
+\frac{7}{4}F_\chi\right)\Gamma^2\ , \label{s1nt1} \\
\label{s2nt2} {\cal S}_2 &=&  e^{-P}\Gamma^4 \partial_\sigma
-\Gamma^5\partial_\delta\ \, . \end{eqnarray} If we compare these
expressions with the supersymmetric constraint in the original
foliation, given in Eqs.(\ref{s1nt}) and (\ref{s2nt}), we see that
the general structure does not change. The integration of the
differential equations following from applying (\ref{s1nt1}) and
(\ref{s2nt2}) to the wave function leads to a solution with a
structure similar to that of Eq.(\ref{wfu1}). This shows that the
general behavior of the wave function of the universe is
qualitatively invariant with respect to a transformation of the
time coordinate as given in Eq.(\ref{ttran}). This coordinate
transformation of time is quite general since it involves all the
nonignorable coordinates present in Gowdy models and implies a
constant lapse function. Clearly, the qualitative independence of
the wave function of the universe holds also in the case of less
general coordinate transformations.

It is necessary to mention that for the expression of the
Lorentz constraint in terms of the gravitino field (\ref{lorentz})
there is only one possible choice, namely
\be
\phi_1 \sim \gamma^1 \ ,\quad
\phi_2 \sim \gamma^2 \ , \quad \phi_3 \sim \gamma^3 \ .
\label{papi1}
\ee
 On the other hand,
for the solution of the supersymmetric constraint we use the choice
(\ref{papi}) which is more convenient for the analysis of the
resulting differential equations. Nevertheless, one could use
the representation (\ref{papi1}) since the final result for
the wave function of the universe does not depend on the particular
used representation.

It is important to emphasize that the physical interpretation of
the wave function of the universe $\vert\Psi\rangle$ presents certain
difficulties.  A
genuine wave function must be related to observable quantities
and this implies that $\vert\Psi\rangle$ must yield a
probability density. However, this is not true in this case,
in particular
because the wave function of the universe is not normalizable.
Moreover, if we require that $\vert\Psi\rangle$ yields a probability
density for the 3-geometry which must have a specific value at a
given time, this would imply a violation of the
Hamiltonian constraint \cite{kucryan}. These difficulties in
the interpretation of the wave function of the universe
are the price one has to pay for applying the canonical
quantization procedure which involves the isolation of a
specific ``time'' parameter against which the evolution
of the system can be defined. An alternative procedure
like the Dirac quantization based on functional integrals,
 which does not require to single
out the time variable, could lead to a quantum system
with less interpretation difficulties \cite{guvryan}.

In this work we have investigated the Gowdy $T^3$ cosmological models
in the context of $N=1$ supergravity. The quantum constraints resulting
from the canonical quantization formalism were explicitly analyzed, and
for the resulting set of differential equations we were able to find
general solutions. In this way, we found the wave function of the
universe for the polarized and unpolarized special cases.
This represents
a proof of the existence of physical states in the $(N=1)$ supersymmetric
midisuperspace corresponding to Gowdy cosmologies. This result contrasts
drastically with analogous investigations in minisuperspace (Bianchi)
models where no physical states exist, a result that  sometimes is
assumed as a sufficient proof to dismiss supergravity $N=1$. We have adopted a
less radical position in this work and dismiss as unphysical only the
minisuperspace models. The existence of physical states in midisuperspace
models confirms this conclusion and indicates that supergravity $N=1$ is a
valuable theory which should be investigated further. In this context
we have also obtained an interesting result showing
that, in the Gowdy $T^3$  midisuperspace model
analyzed in this work,
the wave function of the universe which represents nontrivial physical
states is completely free of anomalies.

Gowdy cosmologies are characterized by the existence of a
curvature singularity at temporal infinity, and it is known that
the metric functions near the singularity evolve according to the
AVTD behavior. We have shown that after the canonical quantization
the corresponding wave function of the universe is free of
singularities. This represents a solution to the singularity
problem and is one of the main results of this work. The mere
presence of the Rarita-Schwinger field and the consideration of a
genuine midisuperspace model is sufficient to eliminate the
classical singularity. This result points to a further interesting
and expected property of supergravity $N=1$ in the sense that it
is able to properly handle the conceptual limits of classical
general relativity.

In this work we focused on the special case of $T^3$ cosmologies. The generalization
of our results to include the case of $S^1\times S^2$ Gowdy models seems to be
straightforward. In particular, we believe that the unified parametrization
introduced in \cite{procmike}, which contains both types of topologies,
could be useful to explore the supersymmetric Gowdy model in quite general
terms.

\begin{acknowledgments}

We would like to thank Michael P. Ryan jr. for stimulating
discussions and literature hints. This work was supported by
CONACyT grants 42191--F, and 36581--E.

\end{acknowledgments}

\end{document}